\documentstyle[prl,aps]{revtex}
\begin{document}
\draft
\twocolumn[\hsize\textwidth\columnwidth\hsize\csname @twocolumnfalse\endcsname

\title {Frequency-dependent spin susceptibility in the
two-dimensional Hubbard model}
\author{
C.E. Creffield, P.E. Kornilovitch,
E.G. Klepfish, E.R. Pike and Sarben Sarkar}
\address{Department of Physics, King's College London, Strand, 
London WC2R 2LS, UK}
\date{\today}
\maketitle
\begin{abstract}
A Quantum Monte Carlo calculation of dynamical spin susceptibility in the 
half-filled 2D Hubbard model is presented for temperature $T=0.2t$ and 
an intermediate on-site
repulsion $U=4t$. Using the singular value
decomposition technique we succeed in analytically continuing 
the 
Matsubara Green's function into the real frequency domain and in deriving the
spectral representation for the longitudinal and transverse spin susceptibility.
The simulation results, while contradicting the random-phase approximation
prediction 
of antiferromagnetic 
long-range order at this temperature, are in agreement with an extension of
a self-consistent
renormalization approach of Moriya. The static susceptibility
calculated using this technique is qualitatively consistent with 
the $\omega \rightarrow 0 $ simulation results.

\end{abstract}
\pacs{PACS numbers: 74.20.-z, 74.20.Mn, 74.25.Dw}
\vskip2pc]
\narrowtext
The dynamical magnetic susceptibility in strongly correlated electron
systems (SCES) has
been at the centre of 
attempts to explain the pairing mechanism in
high temperature superconductivity (HTSC) [\ref{SCHRIEFFER_SDW}],
[\ref{KAMPF}] as well
as to understand the normal-state properties of the HTSC compounds 
[\ref{MORIYA90}],[\ref{MONTOUXPINES}]. The frequency dependence of the magnetic
susceptibility can be measured in neutron scattering experiments, thus allowing 
the assessment of 
the relevance of various low-energy models of SCES to HTSC,
and the validity of their approximate solutions in various temperature
and doping regimes. Quantum Monte Carlo (QMC) simulations, among other
numerical techniques, provide
an alternative test
of these models and of the applicability of the analytical approximations.
These simulations
supply direct information only about the
imaginary-time dependence of the correlation
functions. Therefore, most of the information derived from them
has been concerned with the static susceptibility 
[\ref{SCALAPINO_COMPARISON}]. 
Analytic continuation of 
QMC data into the real-time domain 
has been performed recently to obtain the
single-particle spectral weight function [\ref{WHITE}],
[\ref{SWFpaper}] and two-particle
[\ref{ANDERSONMODEL}] Green's functions. However, a comprehensive 
dynamical description
of the SCES, particularly of their collective excitations, 
based on QMC simulations, remains a largely unexplored area. 
In this letter we report results of analytic continuation by means of
singular value decomposition (SVD) 
to evaluate the
dynamical spin susceptibility. These results allow us to examine spectral
characteristics predicted 
in the spin-density wave (SDW) treatment [\ref{SCHRIEFFER_SDW}]
for
a 2D Hubbard model at finite temperature.

Interpretations of QMC results for magnetic susceptibility have been
mostly related to random-phase approximation (RPA) calculations which lead 
to an SDW ground state [\ref{SCHRIEFFER_SDW}]. 
Comparisons have been made between finite-lattice
simulations and RPA results for an infinite system 
[\ref{SCALAPINO_COMPARISON}],[\ref{MOREOSCALAPINO90}], thus neglecting
the finite-size effects. 
These latter calculations overestimate the value of the 
N\'eel temperature
[\ref{MORIYABOOK}] which would be only enhanced in a calculation done on
a finite lattice [\ref{VEILLEUX}]. Even without this enhancement,
the fit of the QMC data to the RPA prediction
requires an artificial correction to the bare
Coulomb repulsion in the RPA expression for the susceptibility
[\ref{SCALAPINO_COMPARISON}],[\ref{VEILLEUX}],[\ref{BULUT}].
We explored, therefore, an application of
an extended version
of the self-consistent renormalization (SCR) theory which is known
to provide a more accurate description of weak antiferromagnetism
[\ref{MORIYABOOK}],
and found that for a
finite system this theory agrees qualitatively with the QMC results
with no parameter-adjustment.

We apply a finite-temperature QMC algorithm simulating
the partition function of the Hubbard model as a path integral
of a euclidean field theory. The temperature is represented by
an additional compact dimension whose extent is equal to the inverse
temperature $\beta$. Matsubara Green's functions $G_{AB}(\tau)$
for operators $A$ and $B$, satisfying for imaginary-time
separation $\tau$ $G(\tau+\beta)=\pm G(\tau)$
are evaluated as ensemble
averages of the type:
\begin{equation}
G_{AB}(\tau)
={{\int[D\Phi] A(\tau)B(0)\exp(S[\Phi])}\over
{\int[D\Phi] \exp(S[\Phi])}}; \hskip 8pt\,\, (0<\tau\leq\beta )
\end{equation}
where 
$\Phi$ are the generalised
coordinates of the action $S$. 
Analytic continuation of these functions into real time amounts to a 
solution of the following ill-posed inverse problem:
\begin{equation}\label{inverse}
G(\tau)=\int_{-\infty}^\infty{d\omega
{{e^{-\omega\tau}}\over {1\pm e^{-\omega\beta}}}\chi''(\omega)};
\hskip 8pt\,\, (0<\tau\leq\beta ).
\end{equation}
In what follows $\chi''(\omega)$ is the imaginary part of the
commutator retarded Green's
function, which determines the sign in the denominator $(-)$ and the 
periodicity
of the boundary condition.
The solution for $\chi''(\omega)$ 
was found using the singular value decomposition
technique described in Ref. [\ref{SWFpaper}].

We calculate spin-correlation functions
derived from QMC simulation data for Matsubara Green's functions
of the type:
\begin{eqnarray}\label{corf}
&G&_{S^z S^z}(\vec k ,\tau)
\!=\!\langle (n_{\uparrow}(\vec k, \tau)\!-\!n_{\downarrow}(\vec k, \tau)\!)
(n_{\uparrow}(\!-\vec k, 0)\!-\!n_{\downarrow}(\!-\vec k,0 )\!)\rangle 
\nonumber\\
&G&_{S^+ S^-}(\vec k ,\tau)
\!=\!\langle c^\dagger_{\uparrow}(\vec k, \tau)c_{\downarrow}(\vec k, \tau)
\,c^\dagger_{\downarrow}(-\vec k, 0)c_{\uparrow}(-\vec k, 0)\rangle 
\end{eqnarray}
where $c^\dagger_\sigma$ ($c_\sigma$) is an electron 
creation (annihilation) operator with spin projection $\sigma$, 
$n_\sigma=c^\dagger_\sigma c_\sigma$ and $\vec k$ is the lattice momentum.
The simulations were performed for a single-band $2D$ Hubbard model with
on-site Coulomb repulsion $U=4t$ ($t$ being the nearest-neighbour
hopping parameter) at simulation temperature $1/\beta=0.2t$ and lattice-sizes
$8^2$ and $10^2$. The models were simulated at half-filling.
We solve Eq.(\ref{inverse}) with its left-hand-side given by 
$G_{S^z S^z}(\vec k ,\tau)$ and $G_{S^+ S^-}(\vec k ,\tau)$,
its solution $\chi''(\vec k, \omega)$ being the imaginary part
of the dynamical longitudinal ($\chi^{zz}(\vec k, \omega) $) and transverse 
($\chi^{+-}(\vec k, \omega)$)
susceptibilities respectively.

In Fig.1 we present the results for $\chi''(\vec k, \omega)$ for various
values of the lattice momentum $\vec k$. 
The
data is based on 4200 measurements on a $8^2$ lattice and
4100 measurements on a $10^2$ lattice. The ensemble averages and the 
statistical errors were estimated using bootstrap analysis.
There are no qualitative 
differences due to finite-size effects between the two sets
of results, which suggests the possibility of extrapolating the
essential behaviour to the thermodynamical limit at this temperature.
The analytic continuation has been based on reconstruction with seven
singular functions of the kernel in Eq.(\ref{inverse}). This number
corresponds to seven singular values whose ratio to the largest
one is not smaller than the average error (${\cal O} (10^{-2})$). 
The Matsubara Green's
functions were evaluated accordingly at seven values of $\tau$, spaced
correspondinly to the zeros of the seventh singular vector. 

A sum-rule used in Ref.
[\ref{PRELOVSEK}] for the
$t-J$ model can be applied in our case to examine
the ex\-tent of double-occupancy in the given temperature and coupling regime.
With our normalization (see Eq.(\ref{inverse})) this sum-rule reads:
\begin{equation}\label{srule}
{1\over {N}} \int_0^{\infty}
d\omega \sum_{\vec k}{\coth {{\beta\omega}\over 2}\chi ''^{zz}}(\vec k, \omega)
=\left\langle \left( S^z_i \right)^2 \right\rangle 
\end{equation}
where $N$ is the number of the lattice sites.
Our results for the average on-site spin projection 
give $\left\langle \left( S^z_i \right)^2 \right\rangle\approx 0.74$ 
which suggests that
13\% of the sites
are doubly occupied. This violates significantly the constraint assumed
in the t-J model and undermines the validity of extrapolating results
derived for the t-J model to the Hubbard model in this regime of coupling
and temperature [\ref{MOREOTJ}]. 

To compare our results with those derived
from the SDW-RPA treatment we examine our results with respect to the
evidence for long-range antiferromagnetic ordering. 
The  SDW-RPA predicts, for the size of the systems simulated,
a N\'eel temperature
$T_{\rm Neel}^{\rm RPA} \approx 0.7 t$, which is above our simulation
temperature. The same treatment predicts also an SDW gap $\Delta=1.38t$
at our simulation temperature. This gap would result in the vanishing
of the time-ordered
Green's function at the nesting vector $\vec Q = (\pi,\pi)$:
$\chi ''^{zz\hskip 2pt T}(\vec Q,\omega)$ for $\omega < 2\Delta$
[\ref{SCHRIEFFER_SDW}].

We find that 
$\chi''(\vec k,\omega)$ for small values of $\omega$
has a clear peak at $\vec k = \vec Q $ which leads also to a sharply
enhanced
real part of static susceptibility $\chi'(\vec Q,\omega=0)$
calculated using the Kramers-Kronig relations 
(Fig.2). 
However our results
contradict  two essential SDW predictions:
the spontaneous breaking of rotational symmetry and the existence of a
gap in the spectral function for the magnetic correlations.
The approximate preservation of the relation
$\chi^{zz}(\vec k,\omega)=2\chi^{+-}(\vec k,\omega)$ (Fig.1)
indicates that the rotational symmetry is not violated, while the
results for $\chi ''^{zz\hskip 2pt T}(\vec Q,\omega)$ (Fig.3)
show no evidence for the SDW gap. This latter discrepancy with the
SDW prediction cannot be explained by finite-size effects, since the
sign of the gap is invariant under the change of the staggered magnetization.
Therefore if antiferromagnetic
order existed in the thermodynamical limit we would expect 
a minimum (if not vanishing) of 
$\chi''$$ ^{zz\hskip 2pt T}(\vec Q,\omega)$ for small values of $\omega$. This 
is clearly absent in our results. 

We observe therefore that our results cannot be adequately explained within
the SDW-RPA scheme. 
The standard RPA result yields for $T>T_{\rm Neel}$:
\begin{equation}\label{chiRPA}
\chi^{+-}_{\rm RPA}(\vec k,\omega)={{\chi^{+-}_0(\vec k,\omega)}
\over {1-U\chi^{+-}_0(\vec k,\omega)}}.
\end{equation}
We note that the non-interacting susceptibility 
$\chi_0(\vec Q,0)$ calculated for a two-dimensional finite system
at half-filling exhibits a
$1/T$
divergence as $T\rightarrow 0$. 
Therefore, there will always be a pole of 
the susceptibility at some nonzero temperature $T_{\rm Neel}^{\rm RPA}$,
indicated by
$U\chi_0(\vec Q,0)=1$, 
with a subsequent divergence of the total free energy
of a finite system. This unphysical result is one of the consequences
of an inherent
shortcoming of the SDW-RPA treatment, namely, 
its failure to
represent in a self-consistent manner the contribution
of the spin fluctuations to the free energy.
The free energy can be expressed as a functional in the dynamical
susceptibility
[\ref{MORIYABOOK}] (see Eqs.(\ref{freeen}),(\ref{deltaF})).
On substituting expression (\ref{chiRPA}) into this functional and subsequently
differentiating it twice with respect to the staggered magnetization
one fails to recover the static limit of (\ref{chiRPA}). This 
inconsistency has been pointed out [\ref{MORIYABOOK}] as the reason for predicting
too high values for the N\'eel temperature of weak-antiferromagnetic metals.
These arguments 
suggest the need to explore an analytic approach which would
more adequately describe the thermodynamics of the system, if we are
to try to relate our QMC results to any analytical concepts.

Such an approach has been developed as the theory of  
SCR by Moriya et al. [\ref{MORIYABOOK}]. 
A renormalization function $\lambda$ is introduced which relates 
the exact sum of two-particle irreducible diagrams 
$\tilde \chi_{UM}^{+-}(\vec k, \omega)$ to the
free susceptibility $\chi_{0M}^{+-}(\vec k, \omega)$ in a given background
staggered magnetization $M$:
\begin{equation}\label{exact} 
\tilde \chi^{+-}_{UM}(\vec k, \omega)=
{{\chi^{+-}_{0M}(\vec k, \omega)}\over {
1+\lambda_{UM}(\vec k,\omega)}}. 
\end{equation}
This renormalization function enters the calculation
of the full susceptibility:
\begin{eqnarray}\label{fullsusc}
&\chi&^{+-}_{UM}(\vec k,\omega)=
{{\chi^{+-}_{0M}(\vec k, \omega)\Lambda(\vec k+\vec Q)
+U[\chi^{+-}_{0M}(\vec k,\vec k+\vec Q, \omega)]^2}\over
{\Lambda(\vec k)\Lambda(\vec k+\vec Q)-
[U\chi^{+-}_{0M}(\vec k,\vec k+\vec Q, \omega)]^2}}\nonumber\\
&\Lambda&(\vec k)\equiv 1+\lambda_{UM}(\vec k,\omega)-
U\chi^{+-}_{0M}(\vec k, \omega).
\end{eqnarray}
Here the free umklapp susceptibility at staggered magnetization $M$
is denoted by 
$\chi^{+-}_{0M}(\vec k,\vec k+\vec Q, \omega)$.
The free energy of the system with $N_{\rm el}$ electrons is given by:
\begin{equation}\label{freeen}
F(U,M)=
F_0(M)+{U\over N}\left( {{N_{\rm el}^2}\over 4}-M^2\right)+
{\scriptstyle \triangle} F(U,M)
\end{equation}
where the first two terms represent mean-field
free energy and the third one is the correction due to the 
collective excitations which can be expressed 
via $\chi^{+-}_{UM}(\vec k,\omega)$ as: 
\begin{equation}\label{deltaF}
{\scriptstyle \triangle} 
F(U,M)\!=\! -T\!\!\int_0^U\!\! dU'\sum_{\vec k \, \omega_{\nu}}\!\!
\left[ \chi^{+-}_{U'M}(\vec k, i\omega_{\nu}) - \chi^{+-}_{0M}
(\vec k, i\omega_{\nu})
\right]\!\! .
\end{equation}
where $\omega_{\nu}=2\pi T \nu$.
The sum in (\ref{deltaF}) is
dominated by $\chi^{+-}_{UM}(\vec Q,0)$, which justifies 
the long-wave static approximation $\lambda_{UM}(\vec k,\omega)=
\lambda_{UM}(\vec Q,0)$. In this approximation, 
Eqs.(\ref{fullsusc})-(\ref{deltaF}) together with the thermodynamical
relation between the static susceptibility and the free energy:
\begin{equation}\label{selfcons}
{1\over {\chi^{+-}_{UM}(\vec Q,0)}}=
{1\over 2}{{\partial^2 F(U,M)}\over {\partial M^2}}\,_{\displaystyle |_{M=0}}
\end{equation}
constitute a set of equations for the susceptibility,
automatically maintaining the self-consistency requirement.
Eq.(\ref{selfcons})
can be solved by expanding $\lambda_{UM}(\vec Q,0)$ in powers of $M$.
In the next to leading order expansion:
$\lambda_{UM}(\vec Q,0)\approx \lambda_0(U) + (1/2)\lambda_2(U) M^2$
we obtain:
\begin{eqnarray}\label{lambda02}
\lambda_0(U)&=&\chi_{0M}^{+-}(\vec Q,0){{\partial^2 
{\scriptstyle \triangle} F}\over 
{\partial M^2}}\,_{\displaystyle |_{M=0}}
\nonumber\\
\lambda_2(U)&=&{{\partial^2 \chi_{0M}^{+-}(\vec Q,0)}\over {\partial M^2}}
{{\partial^2 {\scriptstyle \triangle} F}\over
{\partial M^2}}\,_{\displaystyle |_{M=0}}.
\end{eqnarray}
Since even
within this approximation the thermodynamics of the spin fluctuations
is accounted for self-consistently, the qualitative description of 
the susceptibility is likely to be 
more accurate than the one provided by SDW-RPA.
Apparently due to the finite-size $1/T$ divergence 
of $\chi_0$,
Moriya's approximation of $\lambda$ independent of $U$ and $M$, used
in the infinite-system calculation [\ref{MORIYABOOK}] is too crude for this
calculation. 
By substituting the solution of
Eq.(\ref{lambda02}) for $\lambda_{UM}(\vec Q,0)$ into
(\ref{exact}) at $\omega=0$ we obtain the value
of $\tilde \chi^{+-}_{U0}(\vec k,0)$. The static susceptibility 
(at $T>T_{\rm Neel}$) is
obtained by replacing $\chi^{+-}_{0}(\vec k,0)$ in Eq.({\ref{chiRPA})
by $\tilde \chi^{+-}_{U0}(\vec k,0)$
(Fig.2). The calculation of the direct and umklapp free
susceptibility entering Eq.(\ref{fullsusc}) was done 
for lattice sizes $8^2$ and $10^2$. 
Thus the finite-size effects, inevitably present
in the QMC simulations, have been accounted for in the analytical
calculation. 
We note the qualitative agreement 
between the calculated values and the simulation results at the 
nesting vector and the rest of the lattice
momenta. The agreement is particularly remarkable since there is no 
parameter adjustment in this analytical calculation. 
Due to the discrete spectrum of single-particle 
excitations on a finite lattice, 
$\chi_0''(\vec k,\omega)$ is a sequence
of $\delta-$functions 
and becomes smooth only for an infinite system. An imaginary
regularizer (corresponding to a finite life-time
for a quasiparticle) has thus to be introduced to allow the calculation
of dynamical susceptibility using Eq.(\ref{fullsusc}). We found
the results to be sensitive to the value of this regularizer, therefore,
a meaningful comparison with the QMC data can be done only for static
susceptibility. 

Making a standard approximation $\chi'(\vec q+\vec Q,0)=\chi'(\vec Q,0) /
(1+\xi^2 \vec q^2) $ for small deviations $\vec q$ from the nesting
vector,
we obtain the correlation length $\xi=3.2 a$ and $\xi=3.6 a$ ($a$ being 
the lattice spacing) 
for the $8^2$ and $10^2$ lattices respectively. 
The SCR estimates are $\xi=3.9 a$ and $\xi=4.4 a$.
We emphasize, however, that
since the correlation length calculated is of the order of the lattice 
linear dimension
this estimate can vary as lattice size increases.

To summarise, the SVD technique is applicable to the derivation of the
dynamical properties of collective
excitations in SCES. 
The dynamical susceptibility obtained from QMC simulations contradicts 
the qualitative predictions of SDW-RPA, in particular there is no evidence for
SDW gap formation in the longitudinal time-ordered correlation
function. Therefore it is unlikely that the gaps observed for this
lattice size and temperature in previous QMC simulations of the 
single-particle density of states
[\ref{WHITE}] can be explained within SDW-RPA as was indeed pointed out in
Ref.[\ref{SWFpaper}]. Our results support the evidence for the 
insufficiency of
the SDW-RPA picture to describe the spectrum of excitations of the Hubbard
model [\ref{FURUKAWA}],[\ref{HUBBARDZEROS}].
Their qualitative
agreement in the static limit with the SCR calculation points to the
importance of including the 
paramagnon interaction in the description of the magnetic
properties of the Hubbard model. This approach has 
previously been used to obtain
on a more phenomenological basis, the dynamics of collective excitations
in HTSC compounds in the normal phase as well as to calculate 
the pairing potential [\ref{MORIYA90}]. Our work outlines a way to
examine these properties directly from the lattice-field
model by means of numerical simulations as well
as by a power series expansion in the magnetisation within the SCR theory.

We are grateful to J.H. Jefferson,
G. Aeppli, A. Bratkovsky and V. Kabanov for stimulating 
discussions. This research was funded
by an SERC grant GR/J18675 and our general development of SVD techniques by
the US Army Research Office, agreement no. DAAL03-92-G-0142. 
Two of the authors (CEC and PEK) were supported by EPSRC and ORS - University
of London respectively.

\noindent{\bf References}
\begin{enumerate}
\item\label{SCHRIEFFER_SDW}{J.R. Schrieffer, X.W. Wen and S.C. Zhang,
{\it Phys. Rev.}{\bf B39}, 11663 (1989);}
\item\label{KAMPF}{A. Kampf and J.R. Schrieffer, {\it Phys. Rev.} {\bf B41},
6399 (1990); {\it ibid.} {\bf B42}, 7967 (1990);}
\item\label{MORIYA90}{T. Moriya, Y. Takahashi and K. Ueda, {\it Jour. Phys. Soc.
 Japan}
{\bf 59}, 2905 (1990);}
\item\label{MONTOUXPINES}{P. Monthoux and D. Pines, {\it Phys. Rev.}
{\bf B49}, 4261 (1994);}
\item\label{SCALAPINO_COMPARISON}{N. Bulut, D.J. Scalapino and S.R. White,
{\it Phys. Rev.}{\bf B47}, 2742 (1993);}
\item\label{WHITE}{S.R. White, {\it Phys. Rev.} {\bf B44}, 4670 (1991);
M. Veki\'c and S.R. White, {\it Phys. Rev.} {\bf B47}, 1160 (1993);}
\item\label{SWFpaper}{C.E. Creffield, E.G. Klepfish, E.R. Pike and Sarben
Sarkar, {\it Phys. Rev. Lett.} {\bf 75}, 517 (1995);}
\item\label{ANDERSONMODEL}{M. Veki\'c {\it et al.,
Phys. Rev. Lett.}{\bf 74}, 2367
(1995); H. Endres {\it et al. , unpublished;}}
\item\label{MOREOSCALAPINO90}{A. Moreo {\it et al.,} {\it Phys. Rev.}
{\bf B41}, 2313 (1990);}
\item\label{MORIYABOOK}{T\^oru Moriya, {\it Spin Fluctuations in
Initerant Electron Magnetism} {\bf Springer-Verlag} (1985);
H. Hasegawa and T\^oru Moriya, {\it Jour. Phys. Soc. Japan} {\bf 36},
1542 (1974);}
\item\label{VEILLEUX}{L. Chen {\it et al.,
Phys. Rev. Lett.} {\bf 66}, 369 (1991);
Alain F. Veilleux {\it et al., unpublished;}}
\item\label{BULUT}{N. Bulut, D.J. Scalapino and S.R. White, {\it Physica}
{\bf C246}, 85 (1995);}
\item\label{PRELOVSEK}{J. Jaklic and P. Prelovsek, {\it Phys. Rev. Lett.} 
{\bf 75},
1340 (1995);}
\item\label{MOREOTJ}{A. Moreo, {\it Phys. Rev.} {\bf B48}, 3380 (1993);}
\item\label{FURUKAWA}{N. Furukawa and M. Imada, {\it J. Phys. Soc. Japan}
{\bf 61}, 3331 (1992);}
\item\label{HUBBARDZEROS}{E.Abraham {\it et al., unpublished;}}
\end{enumerate}
\noindent
{\bf Figure Captions:}
\newline\noindent
Fig.1 $\chi''^{zz}(\vec k, \omega)$ (solid line)
and $\chi''^{+-}(\vec k, \omega)$ (dotted line)
{\it vs.} $\omega$ for selected values
of the lattice momentum given in the brackets in each figure; (a) - lattice
$8^2$, (b) - lattice $10^2$. 
\newline\noindent
Fig.2 Static spin susceptibility {\it vs.} lattice momentum; (a) - lattice
$8^2$, (b) - lattice $10^2$. The results of the SCR calculation are
shown by ($\bullet$) .
\newline\noindent
Fig.3 Time-ordered Green's function 
$\chi''$$^{zz \hskip 2pt T}(\vec Q,\omega)$. Note the absence of
SDW gap.
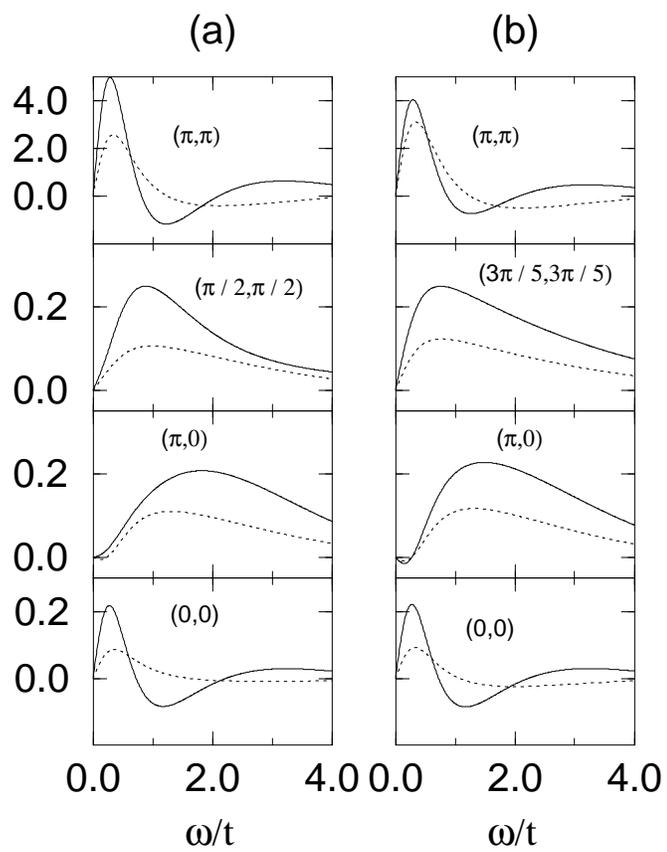
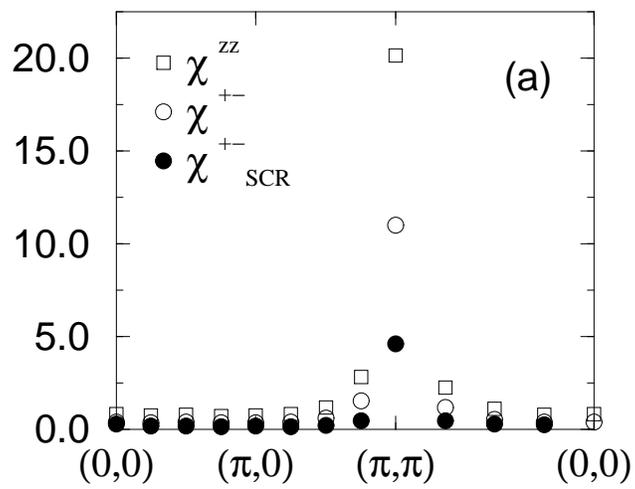
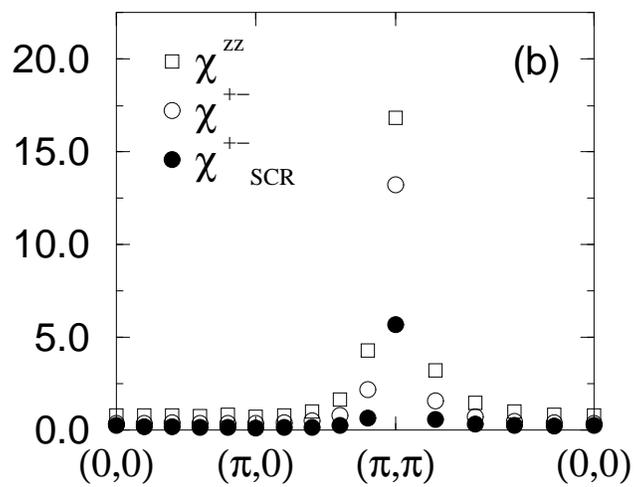
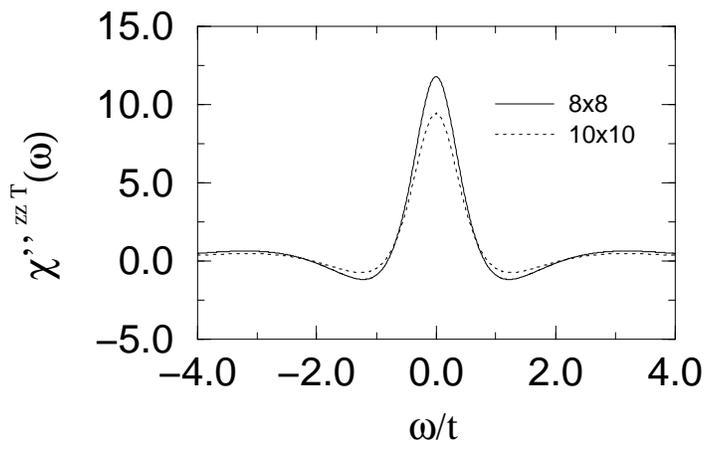
\end{document}